# Thermodynamics of Ion Separation by Electrosorption

Ali Hemmatifar,[a,‡] Ashwin Ramachandran,[b,‡] Kang Liu,[a,c] Diego I. Oyarzun,[a] Martin Z. Bazant,[d] and

Juan G. Santiago[a,*]

‡ contributed equally to this work

[a] *Department of Mechanical Engineering, Stanford University, Stanford, CA 94305, USA*
[b] *Department of Aeronautics & Astronautics, Stanford University, Stanford, CA 94305, USA*
[c] *Wuhan National Laboratory for Optoelectronics, Huazhong University of Science and Technology, Wuhan 430074, China*
[d] *Departments of Chemical Engineering and Mathematics, Massachusetts Institute of Technology, Cambridge, MA 02139, USA*

* To whom correspondence should be addressed. E-mail: juan.santiago@stanford.edu

**Abstract**

We present a simple, top-down approach for the calculation of minimum energy consumption of electrosorptive ion separation using variational form of the (Gibbs) free energy. We focus and expand on the case of electrostatic capacitive deionization (CDI), and the theoretical framework is independent of details of the double-layer charge distribution and is applicable to any thermodynamically consistent model, such as the Gouy-Chapman-Stern (GCS) and modified Donnan (mD) models. We demonstrate that, under certain assumptions, the minimum required electric work energy is indeed equivalent to the free energy of separation. Using the theory, we define the thermodynamic efficiency of CDI. We explore the thermodynamic efficiency of current experimental CDI systems and show that these are currently very low, *less than 1%* for most existing systems. We applied this knowledge and constructed and operated a CDI cell to show that judicious selection of the materials, geometry, and process parameters can be used to *achieve a 9% thermodynamic efficiency* (4.6 kT energy per removed ion). This relatively high value is, to our knowledge, by far the highest thermodynamic efficiency ever demonstrated for CDI. We hypothesize that efficiency can be further improved by further reduction of CDI cell series resistances and optimization of operational parameters.

Ion separation (desalination) processes of any type have long been known to require at least a minimum energy equivalent to the (Gibbs) free energy of separation. The minimum energy consumption corresponds to an ideal, extremely slow reversible process free of resistive losses. In systems with electrostatic interactions, such as supercapacitors, capacitive deionization (CDI), and capacitive mixing (CAPMIX), a common practice to analyze this limit is to start from mechanistic details of electrostatic charge attraction, including specific treatments of the electric double layer (EDL) structure, and then proceed to formulate work integrals to calculate energy.[1–3] A significant limitation of this bottom-up approach is that it cannot be readily generalized to wide range of EDL models, and more importantly, it provides limited insight into the various assumptions and approximations associated with each EDL model. In particular, numerical integration of electrosorptive work generally obscures the effects of compact versus diffuse portions of the EDL structure, effects of finite ion size, and the geometry and relative thickness of EDLs (e.g. whether the EDLs are overlapped or non-overlapped).

In this work, we present an alternative "top-down" theoretical framework that naturally reveals the thermodynamic limits of electrosorption, starting from first principles by variation of the electrochemical free energy functional. Similar variational approaches have been used to calculate electric double-layer surface forces[4–6] and to formulate modified Poisson-Boltzmann models of diffuse charge profiles,[7–11] implicit solvent models[12,13] and phase-field models of ionic liquids and solids.[14–17] We shall see that the variational approach to electrochemical modeling is also convenient to describe the thermodynamic properties of CDI and yield conclusions applicable across the full range of EDL models and chemistries.

One important measure of ion separation in practice is the thermodynamic efficiency of the process ($\eta$), defined as the ratio of free energy of separation to the actual work input. Without a careful system design, energy consumption can be more than an order of magnitude larger than free energy of separation (i.e. $\eta \ll 10\%$). Various sources of irreversibility include ionic resistive and electronical resistive losses which scale with the temporal rate of adsorption/desorption, parasitic losses of Faradaic charge-transfer reactions, dispersion effects and associated flow mixing, and pressure work toward pumping electrolytes. A low resistance and optimized operation design is thus critical for a successful ion separation system.

Although numerous studies focus on increasing thermodynamic efficiency of mature technologies such as reverse osmosis (RO)[18–20] and thermal desalination,[21,22] there have been only a few studies on the thermodynamics of emerging electrosorptive technologies such as CDI. In fact, only a handful of published studies on CDI even report thermodynamic efficiency values.[23,24] Further, there has been no overview of the current typical values of CDI thermodynamic efficiency or of methods by which it can be improved.

We introduce our top-down thermodynamic formulation as follows. Consider a pair of identical porous and electrically conductive electrodes which will be used to trap, transfer, and release ions from two reservoirs of fixed volume: a "diluted" solution and a "brine" reservoir. Each of the two reservoirs is filled with a binary symmetric ($z$: $z$) electrolyte of initial concentration $c_0$ (**Figure 1a**). The volume of the electrode pore space, diluted reservoir, and brine reservoir are respectively $v_p$, $v_D$, and $v_B$. For these, we define recovery ratio as $\gamma = v_D/(v_D + v_B)$. **Figure 1b** (solid lines) shows an example ion separation cycle on $c_{ions}$ vs. $c_\infty$ plane, where we define $c_{ions} = c_+ + c_-$ as total ion concentration trapped by the electrode ($c_\pm$ being pore concentration of cations and anions) and $c_\infty$ as the reservoir bulk concentration. The ion separation cycle can be rationalized as four stages: (1) charging stage, which stores charge in the pores and reduces diluted reservoir concentration to $c_D$, (2) open-circuit exchange of one reservoir for the other, maintaining constant $c_{ions}$ value, (3) discharging stage, which releases salt and increases brine reservoir concentration to $c_B$ and decreases $c_{ions}$ to $c_B$, and (4) an open-circuit return to the first reservoir at constant $c_{ions}$ value. The latter four stages are the same as those considered by Wang et al.[2] in their study of minimum work requirements for CDI. In this construct, mass conservation requires $dc_{ions}/dc_\infty = -v_{bulk}/v_p$, where $v_{bulk} = v_D$ in charging and $v_{bulk} = v_B$ in the discharging phase. We stress that ion separation from a single diluted reservoir to a single brine reservoir (solid lines) is an idealized case useful in rationalizing CDI systems. For example, one can envision a process involving multiple intermediate reservoirs (not just two end points). In an actual CDI system with flow from a common source, charging and discharging is performed relative to some time-varying concentration associated with the effective time-varying ion density in the cell. The dashed curves in **Figure 1b** are shown here as a qualitative representation of a more general case.

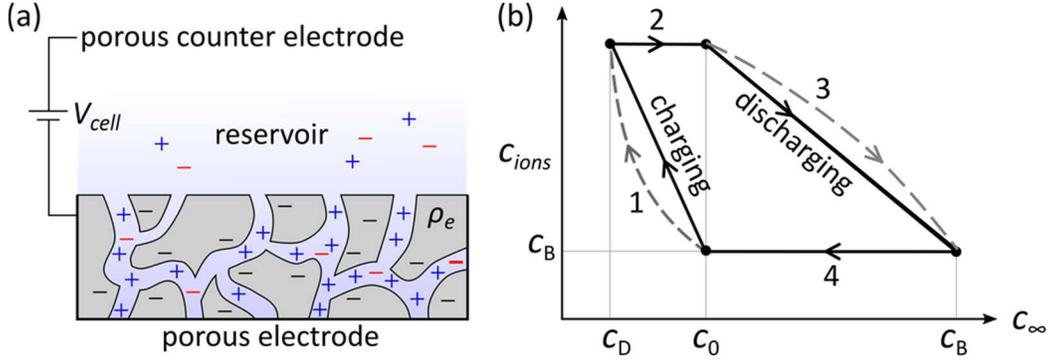

**Figure 1.** (a) Schematic of an electrochemical desalination system with a pair of porous, conductive electrodes (only one shown) in contact with a reservoir. (b) Solid line is an ion separation cycle (on $c_{ions}$ vs. $c_\infty$ plane) wherein charging and discharging are performed with respect to two reservoirs. Dashed lines depict some arbitrary charging/discharging relative to a time-varying value of reservoir concentration.

Our interest is the minimum electrical work required for the ion separation cycle above, which corresponds to the limit of infinitely slow traversal of the charge-voltage cycle with all chemical species in quasi-equilibrium. We seek to formulate this from an energetic perspective without assumptions regarding the details of the charge distribution within the pore electrical double layers (EDLs) or the electrode matrix material. The most general starting point is the total Gibbs free energy functional $G[\{c_i\}, \phi; \rho_e, \phi_e; \{c_j^r\}]$ of a single composite porous electrode, which is an integral over the spatial distributions of ionic concentrations $\{c_i(\vec{x}, t)\}_{i=1}^N$ and the electrostatic potential of mean force $\phi(\vec{x}, t)$ in the electrolyte-filled pore space; the electronic charge density $\rho_e(\vec{x}, t)$ and electrostatic potential (Fermi level per charge) $\phi_e(\vec{x}, t)$ of electrons in the solid electrode matrix; and the concentration profiles $\{c_j(\vec{x}, t)\}_{j=1}^{N_r}$ of any redox-active chemical species (reactants and/or products) that undergo electrochemical reactions with the ions in the pore space. In general, the redox-active species may be charged or uncharged, resulting from Faradaic, auto-ionization or complexation reactions, and may exist anywhere in the porous electrode volume, including dissolved species in the electrolyte filled pores, adsorbed species at interfaces, or inserted species within the electrode solid matrix.

Although our theoretical development below does not depend on the specific form of $G$, it is instructive to point out how various existing mathematical models of electrostatic CDI for flat[25] and porous[26,27] blocking electrodes and Faradaic CDI with redox-active porous electrodes[28–30] are included as special cases our general thermodynamic formalism. In particular, all Poisson-Boltzmann (PB) models (including the classical dilute-solution approximation as well as various modified PB models[9] for thin or thick double layers) correspond to following form of the free energy functional, which expresses the mean-field and local density approximations for a linear dielectric material (neglecting all chemical or electrostatic correlations between ions):[10]

$$G = \int_{V_p} \left( g_c + \rho_p V_T \phi - \frac{\varepsilon}{2} |\nabla \phi|^2 \right) dV_p + \int_{S_e} q_e \phi_e \, dA_e \,, \tag{1}$$

where the integrals are over pore volume and the bounding surface of the solid electrode matrix, respectively; $\rho_p(\vec{x}, t) = \sum_i z_i F c_i$ is the ionic charge density in the electrode pores; $F$ is Faraday's constant (mole of charge); $z_i$ is the valence of ionic species $i$; $g_c(\{c_i\}, T) = h - Ts$ is the chemical portion of the Gibbs free energy density (due to short-range non-electrostatic interactions) expressed in terms of the enthalpy density $h(\{c_i\}, T)$ and the entropy density $s(\{c_i\}, T)$; $q_e(\vec{x}, t)$ is the electronic surface charge density per area of the electrode solid matrix, assumed here to be a metal at constant potential $\phi_e$; and $\varepsilon(\vec{x}, t)$ is the permittivity of the pore solution.

The free energy functional is constructed so that the physically relevant electrostatic potential, which conserves Maxwell's dielectric displacement field, satisfies the stationarity condition of vanishing first variation. This construction leads to Poisson's equation for bulk variations

$$\left( \frac{\delta G}{\delta \phi} \right)_{bulk} = \rho_p + \varepsilon \nabla^2 \phi = 0 \,, \tag{2}$$

and the associated electrostatic boundary condition for surface variations,

$$\left( \frac{\delta G}{\delta \phi} \right)_{surface} = q_e + \hat{n} \cdot \varepsilon \nabla \phi = 0 \,. \tag{3}$$

In the mean-field approximation, the force experienced by each individual ion derives from the self-consistent electric field generated by the mean charge density via Poisson's equation. This approximation is reflected in the form of the free energy functional by defining the electrochemical potential of any species $i$ as its variational derivative with respect to a localized unit concentration perturbation,[16]

$$\mu_i = \frac{\delta G}{\delta c_i} = \frac{\delta g_c}{\delta c_i} + z_i F \phi = \mu_i^\ominus + RT \ln a_i + z_i F \phi \,, \tag{4}$$

where $\mu_i^\ominus$ is the standard reference chemical potential and $a_i$ is the chemical activity of species $i$. By setting each electrochemical potential to a constant value, $\mu_i = \mu_i^\infty$ (for exchange with a reservoir "at infinity"), we define the quasi-equilibrium concentration profiles, $\{c_i^{eq}(\phi, \mu_i^\infty)\}$. Inserting the quasi-equilibrium charge density, $\rho_p^{eq}(\phi, \{\mu_i^\infty\}) = \sum_i z_i F c_i^{eq}(\phi, \mu_i^\infty)$, into Eq. (2) yields a modified Poisson-Boltzmann (PB) equation for concentrated solutions ($a_i \neq c_i$)[9], which reduces to the familiar PB equation in the limit of a dilute solution ($a_i = c_i$):

$$-\varepsilon\nabla^2\phi = \sum_i z_i F c_i^\infty \exp\left(\frac{z_i F(\phi_\infty - \phi)}{RT}\right). \tag{5}$$

Existing models of CDI describe electrosorption of ions in quasi-equilibrium diffuse EDLs via the Gouy-Chapman-Stern (GCS) and modified Donnan (mD) models, which correspond to solutions of the PB equation (Eq. (5)) with the Stern "surface capacitor" boundary condition in the limits of thin and thick double layers, respectively,[26,28–30] and are thus included in our thermodynamic framework. More generally, we could relax the quasi-equilibrium assumption in our thermodynamic formulation by defining mass fluxes in terms gradients of the electrochemical potentials defined by Eq. (4) in order to obtain modified Poisson-Nernst-Planck equations for diffuse-charge dynamics out of equilibrium,[9,13,16] which capture additional effects of non-equilibrium charging and tangential surface conduction.[31] Since such non-equilibrium phenomena introduce additional resistance into the system, however, we shall focus on quasi-equilibrium electrosorption, in order to derive a thermodynamic bound on the consumed energy.

Let us define the specific free energy per volume of the diluted reservoir, $g = G/v_D$, and assume macroscopic quasi-equilibrium across the electrode pores. The electrochemical potential $\mu_i = v_D \frac{\delta g}{\delta c_i} = \mu_i^\ominus + RT(\ln a_i + z_i \phi)$, is then *uniform* and *constant* for each ionic species $i$, where $\mu_i^\ominus + RT \ln a_i = \partial g_c / \partial c_i$ is the familiar chemical potential and $a_i$ the activity of species $i$. In the case of dilute electrolyte, $a_i = c_i$. At this point, we can calculate the electrostatic work of the system for the complete ion separation cycle of **Figure 1b**. Assume a differential electronic charge $\delta q_e = \delta \rho_e dV_e$ is injected to the electrodes at fixed electrode potential $\phi_e$, under quasi-equilibrium ($\mu_i$ uniform and constant) exchange with bulk reservoir solution, and results in a change in pore ion concentration of $\delta c_i(\vec{x})$. This process requires a minimum electrostatic work of $-2V_T \tilde{\phi}_e \delta q_e$, where $\tilde{\phi}_e = \phi_e/V_T$ and $V_T$ is the thermal voltage. Due to electroneutrality, ionic charge compensates for electronic charge as in $\delta\rho_p \, dV_p + \delta\rho_e \, dV_e = 0$. Without loss of generality, we take $\phi = 0$. The change in free energy of the electrode for this process is then $v_D \delta g = \left[\int_{V_p} \sum_i \mu_i \, \delta c_i \, dV_p + \int_{V_e} V_T \, \delta\rho_e \, \tilde{\phi}_e \, dV_e\right]$ and using electroneutrality condition we further simplify to

$$v_D \delta g = \sum_i (\mu_i - z_i RT \tilde{\phi}_e) dN_i, \tag{6}$$

with $dN_i = \int_{V_p} \delta c_i dV_p$. For a closed loop, we have $\oint \delta g = 0$, and thus the minimum volumetric energy consumption for the cycle $E_{v,min}$ is

$$E_{v,min} = -\frac{V_T}{v_D}\oint 2\tilde{\phi}_e dq_e = \frac{2}{v_D}\oint \sum_i \mu_i dN_i = \Delta g_{sep}, \tag{7}$$

where $\Delta g_{sep}$ is specific Gibbs free energy of separation. Refer to Supporting Information (SI) for details. Equation (7) states that in the absence of chemical (non-electrostatic) portion of the free energy, $E_{v,min}$ is equivalent to the Gibbs free energy of separation $\Delta g_{sep}$ for any thermodynamically-consistent electrical double layer (EDL) of arbitrary geometry and thickness (overlapped or non-overlapped). This includes Poisson-Boltzmann (PB) type EDL treatments such as Gouy-Chapman (GC) model as well as modified Poisson-Boltzmann with finite ion size effects.[8,32] EDL descriptions with constant or $c_i$-dependent non-electrostatic potentials such as modified-Donnan (mD) model[27,28] each also respect Eq. (7) (since the integral of the non-electrostatic term $\mu_{att}$ on a closed path vanishes). Importantly, as noted above, Eq. (7) is also valid for EDLs with compact layers such as Gouy-Chapman-Stern (GCS) and Gouy-Chapman-Stern-Carnahan-Starling (GCS-CS) models. These conventional electrosorptive desalination EDL models satisfy the free energy formulation results in general, since these models can be derived variationally from the same free energy model.

Equation (7) can be further simplified for 1:1 dilute electrolytes ($a_i = c_i$) and assuming an idealized instantaneous, quasi-equilibrium ion exchange between reservoir and the pores. Under these conditions and for $\phi = 0$ in the reservoir, $\mu_i \approx \mu_{i,\infty} = RT \ln c_\infty$. Moreover, $2dN_i = -v_{bulk}dc_\infty$ by mass conservation. The minimum volumetric electrical work then takes the form

$$E_{v,min} = \frac{2RT}{v_D}\oint v_{bulk} \ln c_\infty \, dc_\infty. \tag{8}$$

For the ion separation cycle of **Figure 1b** (solid lines), $E_{v,min}$ can be written in the familiar form $2RT\left[c_D \ln c_D + \left(\frac{1}{\gamma} - 1\right)c_B \ln c_B - \frac{c_0}{\gamma}\ln c_0\right]$.

Alternative to the approach above, for special cases of non-overlapping (e.g. GCS) and highly-overlapped (e.g. mD) double layer models, the equivalency of $E_{v,min}$ and $\Delta g_{sep}$ can be shown by explicit calculations (direct integration of electrical work). For example, consider a model including a Stern layer assumption. For the equilibrium condition, the (non-dimensional) potential of a single electrode can be written as $\tilde{\phi}_e = \Delta\tilde{\phi}_{St} + \Delta\tilde{\phi}_D$ where $\Delta\tilde{\phi}_{St}$ is the Stern layer potential drop and $\Delta\tilde{\phi}_D$ is either the diffuse layer (in GCS model) or Donnan potential drop (in mD model), both normalized by the thermal voltage. A close inspection of the two models reveals that they each obey the following relation between pore ionic concentration $c_{ions}$ and pore charge density $\rho_p$ for 1:1 electrolytes

$$\frac{1}{F}\frac{\rho_p}{c_\infty^n} = \frac{d}{d\Delta\tilde{\phi}_D}\left(\frac{c_{ions}}{nc_\infty^n}\right), \quad (9)$$

where $n = 1/2$ and $1$ for GCS and mD models, respectively. Using this ad-hoc relation, the electroneutrality condition ($q_e + v_p\rho_p = 0$), and mass conservation ($dc_{ions}/dc_\infty = -v_{bulk}/v_p$), the electrical work can be written as (see SI for details)

$$E_{v,min} = 2\frac{V_T}{v_D}\left[-\underbrace{\oint \Delta\tilde{\phi}_{St} dq_e}_{=0} - \underbrace{\oint d[q_e\Delta\tilde{\phi}_D + Fv_p(1/n - \ln c_\infty)c_{ions}]}_{=0} + F\oint v_{bulk} \ln c_\infty\, dc_\infty\right], \quad (10)$$

or again simply $E_{v,min} = \Delta g_{sep}$. Importantly, note the electrical work of the Stern layer (the first term on the right-hand side) vanishes on a closed loop for the classical surface-capacitor approximation introduced by Grahame,[33] where $\Delta\tilde{\phi}_{St}$ is solely a function of surface charge density $q_e$ (equal to the total nearby pore charge density by macroscopic electroneutrality). We can thus write the integrand as a total differential $dE_{St}$, where $\Delta\tilde{\phi}_{St} = dE_{St}/dq_e$, whose integral around a closed loop must vanish. The second integral must also vanish for any closed loop. In the classical picture, the Stern layer charge and discharge dissipates no energy in a cycle, since its capacitance depends only on charge, and not voltage. In contrast, diffuse layer charging cycles consume energy, since EDL capacitance is inherently voltage dependent, except in the limit of small EDL voltage drop (Debye-Hückel theory), where linear response again yields a constant capacitance, independent of voltage.

Although the preceding analysis is based on the electrostatic free energy functional (Eq. (1)) for electrosorption of ions in the diffuse double layers of a porous electrode, it also applies to CDI processes involving chemical or electrochemical reactions. A simple extension would include regulation of surface chemical charge of the porous electrode, e.g., by pH-dependent deprotonation or hydroxyl dissociation reactions.[34,35] A broader class of reactive systems include Faradaic CDI, where significant electrosorption is mediated by reversible electron transfer reactions that reduce or oxidize the adsorbed ions, either to form new molecules or ions in the pores or on the solid surface[14,29,30] or to intercalate ions into an active solid matrix of the porous electrode.[36–42]

In order to describe such situations, an additional term, $G_F = \int_{V_F} g_F(\{c_j\}, \phi, \phi_e) dV_F$, must be added to the electrostatic free energy functional (Eq. (1)), where the integral is over the "reactive volume" containing all the active species which undergo chemical or electrochemical reactions involving ions in the pores, which have concentration profiles $\{c_j(\vec{x}, t)\}$, homogeneous free energy density $g_F$, and chemical potentials $\mu_j(\vec{x}, t) = \frac{\delta G_F}{\delta c_j}$. In the case of Faradaic reactions with different redox species in each electrode, the open circuit voltage is generally non-zero, and the transient voltage under general non-equilibrium conditions is determined by a stoichiometric sum of the chemical potentials of the reactive ions and reduced/oxidized species, which are defined variationally from the total free energy functional according to the general thermodynamic framework of electrochemical kinetics.[16]

As with electrostatic adsorption in the diffuse EDLs described above, the free energy change of any charge/discharge cycle is minimized when Faraday reactions are also in quasi-equilibrium. In the case of chemical surface charge, a suitable adsorption isotherm governs the thermodynamics. For Faradaic reactions, the quasi-equilibrium charge-voltage relation is derived from a generalized form of the Nernst equation,[16] such as the Frumkin isotherm (or regular solution model) for ion intercalation.[42] In this limit of "fast" reactions, there is no net change in total chemical potential from reactants to products, and the total free energy is the same as in the original electrostatic adsorption model, at the same state of charge. Moreover, since the additional contribution to the energy only depends on the state of charge under the assumption of quasi-equilibrium reactions, it must also vanish under any closed charging cycle, $\Delta G_F = v_D \oint \delta g_F = 0$. Therefore, our main result, Eq. (7), is unchanged by the presence of any chemical or electrochemical reactions in the electrodes.

Having established a general theory of minimum energy consumption during CDI cycles, we now focus on practical relevance of this analysis. First, we highlight the importance of series resistance (an important cell design parameter and fabrication challenge), and then we focus on effects of flowrate and ion separation rate (operational parameters) on thermodynamic efficiency. Note, in CDI, the (ionic and electronic) resistances and Faradaic reactions are two main sources of energy loss.[43,44] The pumping energy is usually insignificant,[43] except perhaps for flow-through CDI designs at very high flow rates. In its simplest form, the major contributor to volumetric resistive consumption $E_{v,resist}$ can be estimated as $R_s I^2/\gamma Q$ where $R_s$, $I$, and $Q$ are respectively the series resistance, current magnitude, and flow rate. To put the energetic performance of current CDI technology into perspective, we show in **Figure 2** experimental performance data for variety of reported CDI cells. The figure shows experimental volumetric energy consumption $E_v$ versus $E_{v,resist}$ for various CDI and membrane-CDI systems under constant-current (CC) operation.[43–49] In addition to published data, we include measurements from a cell which we constructed specifically for the current study. Our CDI cell here features a low-resistance (for each electrode pair) design which we realized by pressing into and embedding titanium mesh current collectors into 270 μm thick commercially available porous carbon electrodes. We also lower series resistance by using 30 μm spacers. The feed solution was 20 mM NaCl feed solution. The cell is a simple flow-between type design with no membranes (see SI for details). We here focus on CC operation as the energy recovery during discharge step results in a lower energy per removed ion compared to short-circuit discharge at 0 V.[23] Despite the wide variation in CDI cell materials, geometry, and operational parameters (including applied current, maximum and minimum voltage settings, and flow rate), the data organize fairly well near and just above the line set by the resistive limit $E_v = E_{v,resist}$ line. This shows our simple estimate of $E_{v,resist}$ correctly predicts an approximate minimum for $E_v$ in practice. Further, it confirms that work towards decreasing series resistance is essential to the improvement of thermodynamic efficiency $\eta$ of CDI cells. Also, we note that for cells with low series resistance, the Faradaic losses can contribute to most of the energy loss. So, remedies for reduction of Faradaic losses can also greatly improve the energy performance of CDI.

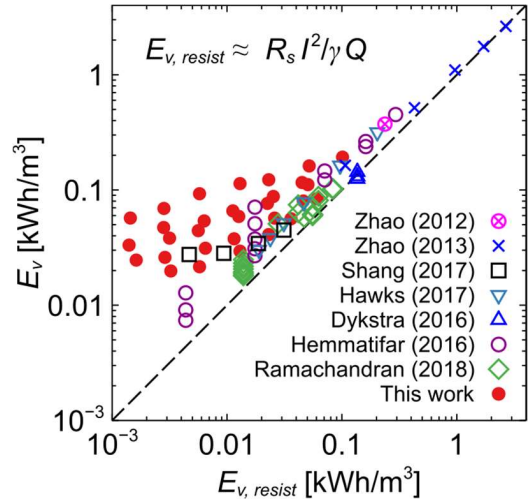

**Figure 2.** Measured volumetric energy consumption ($E_v$) versus estimated resistive energy consumption ($E_{v,resist}$) for a wide variety of CDI cell designs operated under constant-current (CC) operation. $E_{v,resist}$ provides a very good estimate of minimum $E_v$ observed in practice.

Consistent with the importance of series resistance, slow ion separation processes are often more thermodynamically efficient than faster processes. However, slower processes are not always superior because of the effects of Faraday losses. We explore and demonstrate this in **Figure 3**. Shown is data given our low-resistance CDI cell for CC operation with currents in 2.5-65 mA and flow rates in 0.22-3.47 ml/min range. **Figure 3a**, **3b**, and **3c** are respectively the volumetric energy consumption $E_v$, energy consumption per ion and per salt removed (note two ordinates), and thermodynamic efficiency $\eta$, all plotted versus applied current. The voltage windows were 0.4-1 V to ensure high charge efficiency.[49,50] For a given flow rate, operation at high currents is limited by resistive losses as expected.[44] However, at overly low applied current, performance is limited by Faradaic losses since the cell spends overly long times near the high voltage limit.[44] This trade-off between two very different energy loss mechanisms results in pronounced optima in $\eta$. Similarly, exceedingly low flow rate operation (e.g. 0.22 ml/min) suffers from dispersion and mixing inefficiencies associated with incomplete removal of desalted water or brine during charging and discharging, respectively.[51] Instead of a monotonic trend, we see that mid-range values of current and flow rate (10 mA and 0.44 ml/min in this case) results in thermodynamic efficiency of about 9% and energy consumption of about 4.6 kT/ion. This thermodynamic efficiency is unprecedentedly high for traditional CDI designs (i.e. with no membranes, no ion intercalation, no inverted operation, etc.). We again note that such high thermodynamic efficiency is a result of low resistances and careful choice of operation parameters, such as voltage window, current, and flow rate. We predict that even higher $\eta$ can be achieved using optimized operation and high-performance electrodes with low electrical resistance.

Lastly, for completeness, we stress that the energy-efficient ion separation of our cell comes at a price. Any practical operational design is a trade-off between throughput and energy efficiency. **Figure 4a** shows thermodynamic efficiency of ion separation versus desalination depth ($\Delta c$) for our cell and published CC cell data. **Figure 4b** shows the main trade off for this thermodynamic performance is a mediocre value for productivity (defined as volume of treated solution per unit time per cell area)[49] Clearly, for the same range of $\Delta c$, one can achieve significantly higher $\eta$ by sacrificing productivity (low flow rates).

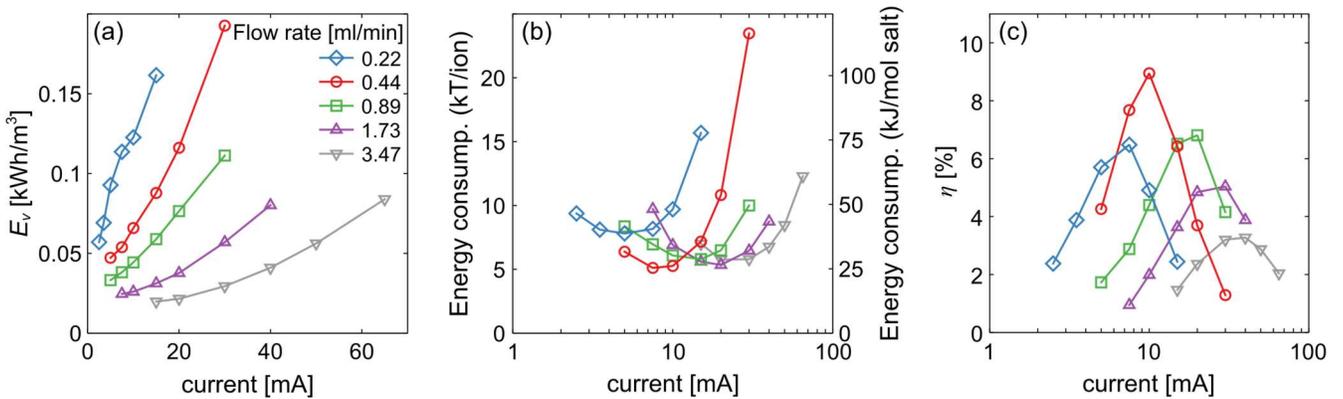

**Figure 3.** (a) Measurements of volumetric energy consumption $E_v$, (b) energy consumption per ion and salt removed, and (c) the thermodynamic efficiency $\eta$ for CC operation at various currents and flow rates. Mid-range values of current (10 mA) and flow rate (0.44 ml/min) results in thermodynamic efficiency of up to 9%.

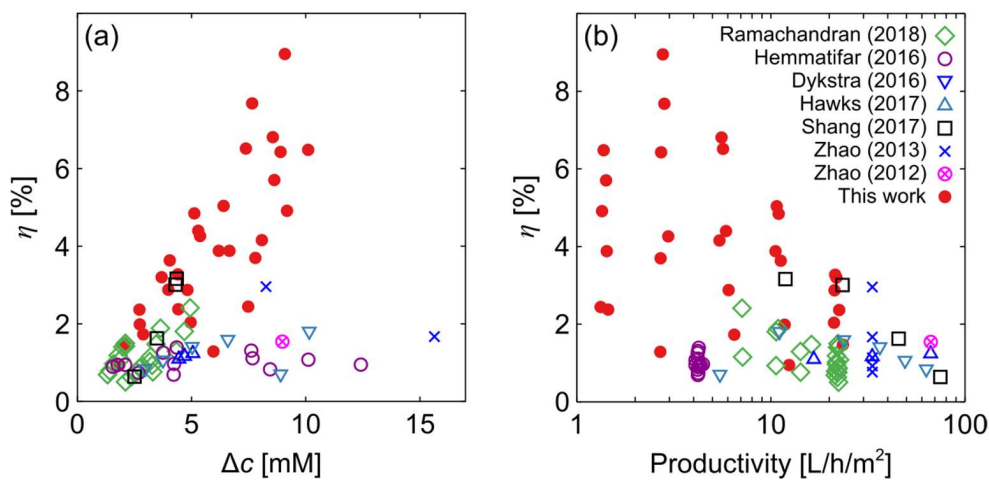

**Figure 4.** Thermodynamic efficiency versus (a) average effluent concentration reduction Δ$c$ and (b) productivity for CDI systems of **Figure 2**. High thermodynamic efficiency generally correlates with low throughout productivity.

To summarize, we presented a top-down approach to calculate minimum energy requirement of electrosorptive ion separation using variational form of free energy formulation. The advantage of our formulation is its generality, free of any assumptions about the charge distribution in the double layers or the profile of Faradaic reaction products, since all such processes occur in quasi-equilibrium (without any internal resistance to transport or reactions) in order to achieve minimum energy consumption. We showed, analytically, the minimum energy required for most known EDLs (irrespective of EDL geometry or thickness) is indeed equivalent to free energy of separation. These include (but not limited to) Poisson-Boltzmann type models such as GCS, GCS-CS, and mD. We focused our attention of the thermodynamic efficiency $\eta$ of CDI systems as an example of electrosorptive ions separation method. We then discussed practical considerations in CDI cell design and operation, and showed cell series resistance and optimization of operational parameters such as flow rate and current are two contributing factors to practical values of $\eta$. As a comparison to published studies, we fabricated and tested a cell with reduced cell resistance. We used variations in operation to identify Faraday losses, resistive losses, and flow efficiency losses as primary irreversible loss mechanisms. Our cell demonstrated an unprecedented 9% thermodynamic efficiency and only 4.6 kT energy requirement per removed ion (NaCl solution), but we show that this achievement comes at the cost of productivity. Overall, the study identifies fundamental limits for CDI operation as well as the significant challenges associated with approaching this limit.

**Acknowledgements**

A.H. gratefully acknowledges the support from the Stanford Graduate Fellowship program of Stanford University. A.R. gratefully acknowledges the support from the Bio-X Bowes Fellowship of Stanford University. We acknowledge partial support from the California Energy Commission grant ECP-16-014, and thank Dr. Michael Stadermann from Lawrence Livermore National Laboratory for fruitful discussions.

**Supporting Information**

The Supporting Information presents details of derivation, details of our cell fabrication and characterization, and supplementary measurements of cell performance.

# Thermodynamics of Ion Separation by Electrosorption

Ali Hemmatifar,[a,‡] Ashwin Ramachandran,[b,‡] Kang Liu,[a,c] Diego I. Oyarzun,[a] Martin Z. Bazant,[d] and Juan G. Santiago[a,*]

‡ contributed equally to this work

[a] *Department of Mechanical Engineering, Stanford University, Stanford, CA 94305, USA*
[b] *Department of Aeronautics & Astronautics, Stanford University, Stanford, CA 94305, USA*
[c] *Wuhan National Laboratory for Optoelectronics, Huazhong University of Science and Technology, Wuhan 430074, China*
[d] *Department of Chemical Engineering and Mathematics, Massachusetts Institute of Technology, Cambridge, MA 02139, USA*

**Abstract**

This document includes details of energy derivations, CDI cell design, cyclic voltammetry and electrochemical impedance spectroscopy tests, constant-current experiment results, and salt adsorption and energetic performance of our CDI cell.

**S.1 Equivalency of minimum electrical work and free energy of separation**

Details of derivation of minimum electrical work for ion separation of Figure 1b of the main text are discussed here. Assume electronic charge $\delta q_e = \delta\rho_e dV_e$ is injected to the electrode at fixed electrode potential $\tilde{\phi}_e$ (normalized by thermal voltage $V_T$). Also, assume this process changes the pore ion centration by $\delta c_i$ according to a quasi-equilibrium ion exchange between reservoir and the pore electric double layers (EDLs). The chemical potential of species $i$, $\mu_i$, thus remains constant. By definition, the work associated with injection of charge $\delta q_e$ is $-V_T \tilde{\phi}_e \delta q_e$. Since $\tilde{\phi}_e$ remains constant and $\phi = 0$, the change in free energy of the electrode is

$$\delta g = \frac{1}{v_D}\int_{V_p} \delta g_c \, dV_p + \frac{V_T}{v_D}\int_{V_e} \delta\rho_e \tilde{\phi}_e \, dV_e, \tag{S.1}$$

where $\delta g_c = \sum_i \mu_i \delta c_i$ by definition. Using electroneutrality condition, $\delta\rho_p \, dV_p + \delta\rho_e \, dV_e = 0$, and the definition $\rho_p = \sum_i z_i F c_i$, Equation S.1 can simplified as (note $V_T = RT/F$)

$$\delta g = \frac{1}{v_D}\int_{V_p} \left[\sum_i (\mu_i - z_i RT \tilde{\phi}_e)\delta c_i\right] dV_p. \tag{S.2}$$

Since $\mu_i$ and $\phi_e$ are constant and uniform throughout the electrode pore volume, we switch the order of summation and volume integral to arrive at Equation 6 of the main text

---

* To whom correspondence should be addressed. E-mail: juan.santiago@stanford.edu

$$\delta g = \frac{1}{v_D} \sum_i \left[ (\mu_i - z_i RT \tilde{\phi}_e) \int_{V_p} \delta c_i dV_p \right], \tag{S.3}$$

with $dN_i = \int_{V_p} \delta c_i dV_p$. So, on a closed loop with $\oint \delta g = 0$ and

$$E_{v,min} = \frac{1}{v_D} \oint \sum_i \mu_i dN_i = \Delta g_{sep}. \tag{S.4}$$

## S.2  Direct integration of minimum electrical work

As an alternative to general approach of the main text, we here prove $E_{v,min} = \Delta g_{sep}$ in special case of Gouy-Chapman-Stern (GCS) and modified Donnan (mD) models (as examples of non-overlapping and highly-overlapped electrical double layers) by explicitly calculation of electrical work. Under equilibrium conditions in absence of electric currents, the electrode potential (normalized by thermal voltage $V_T$) is $\tilde{\phi}_e = \Delta\tilde{\phi}_{St} + \Delta\tilde{\phi}_D$. Pore ionic concentration $c_{ions}$ and pore charge density $\rho_p$ in GCS and mD models can be written as

$$\begin{cases} \rho_p = 4\lambda_D F c_\infty \sinh(\Delta\tilde{\phi}_D/2) \\ c_{ions} = 4\lambda_D c_\infty \cosh(\Delta\tilde{\phi}_D/2) \end{cases} \quad \text{GCS model} \tag{S.5}$$

$$\begin{cases} \rho_p = 2F c_\infty \exp(\mu_{att}) \sinh(\Delta\tilde{\phi}_D) \\ c_{ions} = 2 c_\infty \exp(\mu_{att}) \cosh(\Delta\tilde{\phi}_D) \end{cases} \quad \text{mD model} \tag{S.6}$$

where $\lambda_D$ is Debye length and $\mu_{att}$ is attraction parameter which accounts for non-electrostatic ion adsorption. One can relate $c_{ions}$ and $\rho_p$ through the single equation

$$\frac{1}{F} \frac{\rho_p}{c_\infty^n} = \frac{d}{d\Delta\tilde{\phi}_D} \left( \frac{c_{ions}}{nc_\infty^n} \right), \quad n = \begin{cases} 1/2 & \text{GCS model} \\ 1 & \text{mD model} \end{cases} \tag{S.7}$$

such that

$$\rho_p d(\Delta\tilde{\phi}_D) = F[(1/n) dc_{ions} - c_{ions} d\ln c_\infty], \tag{S.8}$$

or equivalently,

$$\rho_p d(\Delta\tilde{\phi}_D) = F[d((1/n - \ln c_\infty) c_{ions}) + \ln c_\infty \, dc_{ions}], \tag{S.9}$$

and using mass conservation $dc_{ions}/dc_\infty = -v_{bulk}/v_p$,

$$\rho_p d(\Delta\tilde{\phi}_D) = F \left[ d((1/n - \ln c_\infty) c_{ions}) - \frac{v_{bulk}}{v_p} \ln c_\infty \, dc_\infty \right]. \tag{S.10}$$

Now, the differential electrical work per unit desalted volume is $-2 \frac{V_T}{v_D} \tilde{\phi}_e dq_e$ can be expanded as

$$dE_{v,min} = -2 \frac{V_T}{v_D} (\Delta\tilde{\phi}_{St} + \Delta\tilde{\phi}_D) dq_e = -2 \frac{V_T}{v_D} \left( \Delta\tilde{\phi}_{St} dq_e + d(q_e \Delta\tilde{\phi}_D) - q_e d(\Delta\tilde{\phi}_D) \right). \tag{S.11}$$

Electrical work on a closed loop can be derived by substituting Equation S.10 into S.11 and use of electroneutrality condition $q_e + v_p \rho_p = 0$

$$E_{v,min} = -2 \frac{V_T}{v_D} \left( \oint \Delta\tilde{\phi}_{St} dq_e + \oint d\left[ q_e \Delta\tilde{\phi}_D + v_p F \left( \frac{1}{n} - \ln c_\infty \right) c_{ions} \right] - F \oint v_{bulk} \ln c_\infty \, dc_\infty \right). \tag{S.12}$$

In the absence of native surface charges, the Stern potential relates to electronic charge density $q_e$ as $q_e = C_{St} \Delta\tilde{\phi}_{St} V_T$, where $C_{St}$ is Stern layer capacitance. So, assuming a constant or $q_e$-dependent Stern capacitance, Stern layer does not contribute to electrical work (first integral on the right-hand side vanishes) and thus does not consumes energy. The second term also vanishes by definition and we recover $E_{v,min} = \Delta g_{sep}$ for the case of dilute 1:1 electrolyte as

$$E_{v,min} = \frac{2RT}{v_D} \oint v_{bulk} \ln c_\infty \, dc_\infty . \tag{S.13}$$

## S.3 CDI cell design and characterization

The cell assembly was similar to that in [1] with a radial flow structure. Two pairs of square shaped activated carbon electrodes (Materials & Methods, PACMM 203, Irvine, CA) with 5×5 cm size and initial (uncompressed) thickness of 270 μm and total dry mass of 1.14 g was used. We ultrasonically-cleaned titanium meshes of 250 μm thickness (uncompressed) and used them as current collectors. Each current collector was square shaped with 5×5 cm size and had a tab section (1×3 cm) for connection to external wires. To enhance electrical contact between current collectors and electrodes, we embedded the titanium meshes intro the electrodes using a hot press at 100 ºC temperature. A schematic of a single current collector with compressed thickness of 170 μm embedded into carbon electrode is shown in Figure 1a. The compressed thickness of the stack was 300 μm. Electrode side, titanium mesh side, and cross-section views are shown in Figure 1b to 1d, respectively. Polypropylene meshes of 30 μm thickness (McMaster-Carr, Los Angeles, CA) acted as spacers between the electrode pairs. The assembly was then sealed with gaskets and fastened by C-clamps.

We further characterized the cell assembly by cyclic voltammetry (CV) and electrochemical impedance spectroscopy (EIS) tests. The CV test in Figure 2a was performed at 0.2 mV/s scan rate with 20 mM NaCl solution and showed cell capacitance of about 17 F. EIS test in Figure 2b shows 0.33 Ω setup resistance (ionic resistance of the solution in the separators, electrical resistance of current collectors, and resistance of connecting wires). The contact resistance, usually associated with a semi-circle feature, is almost completely eliminated, as shown by inset of Figure 2b.

## S.4 Experimental procedure

The experimental setup consisted of our fbCDI cell, a 3 L reservoir filled with 20 mM NaCl solution, a peristaltic pump (Watson Marlow 120U/DV, Falmouth, Cornwall, UK), a sourcemeter (Keithley 2400, Cleveland, OH), and a flow-through conductivity sensor (eDAQ, Denistone East, Australia). We operated our cell at constant current (CC) charging and discharging and studied the effect of flow rate and current on the thermodynamic efficiency of the CDI desalination process. We used flow rates varying between 0.2-3.5 mL/min, and current values between $I = 5$ to 65 mA with closed-loop circulation in all of our experiments (flow from reservoir to cell and back to reservoir). For all experiments, we charged (at current $I$) and discharged (at current $-I$) the cell between 0.4 to 1.0 V voltage window. Further, we performed at least three complete charge/discharge cycles for each experiment to ensured dynamic steady state (DSS) operation, in which salt adsorption during charging is equal to desorption during discharging. We recorded external voltage and effluent conductivity using a Keithley sourcemeter and an eDAQ conductivity sensor (with ~93 μL internal channel volume). Conductivity was converted to salt concentration using a calibration curve for NaCl. In Section S.5, we present voltage and concentration measurements versus time for our experiments.

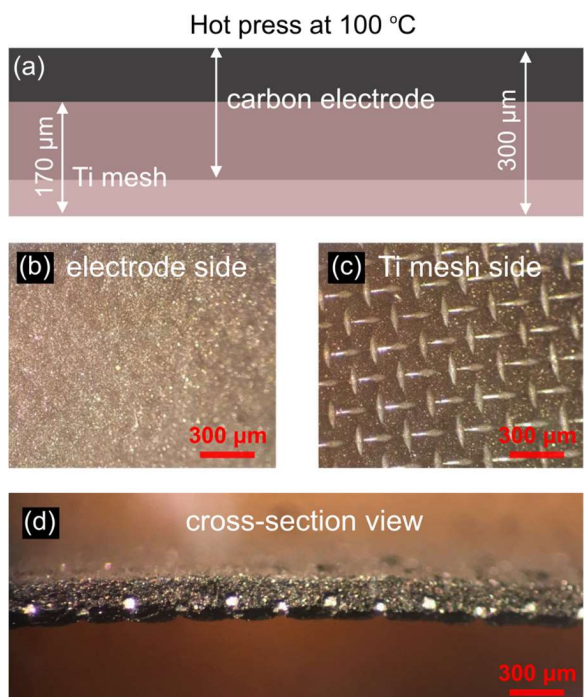

Figure 1. (a) Schematic of embedded titanium mesh sheet into the acticated carbon electorde using 100 ºC hot press. Images of (b) electrode side, (c) titanium mesh side, and (d) cross-section side views after hot press step.

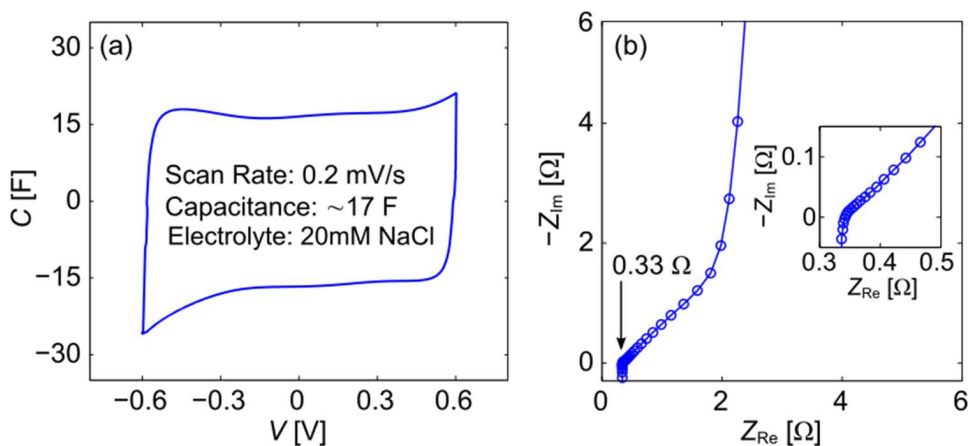

Figure 2. (a) Cyclic voltammetry (CV) test of the assembled cell at 0.2 mV/s scan rate and 20 mM NaCl electrolyte solution with -0.6 to 0.6 V voltage window. CV test shows about 17 F cell capacitance. (b) Electrochemical impedance spectroscopy (EIS) test of the assembled cell again with 20 mM NaCl electrolyte solution. The setup resistance (ionic resistance in separators, electric resistance of current collectors and wires) is only 0.33 Ω and the contact resistance is negligible.

## S.5 Voltage and concentration profiles

Figure 3 shows selected voltage and effluent concentration profiles for constant current charging and discharging of our CDI cell with 20 mM NaCl inlet salt concentration. First, second, and third column respectively correspond to operation at 5, 10, and 20 mA currents. Labels in each panel show the flow rate used in ml/min. All data correspond to dynamic-steady-state (DSS) condition wherein concentration and voltage profiles do not vary between the cycles. DSS reached after 3 to 5 cycles depending on the current and flow rate values. Voltage window for all the experiments was 0.4-1 V to achieve higher charge efficiency compared to 0-1 V voltage window.[2] For experiments at fixed charge/discharge current (each column), the value of flow rate dramatically affects both voltage and concentration profiles. Operation at low flow rates leads to diffusion-limited regime where mass transport of ions is the major limiting factor.[3] So, the cell undergoes salt depletion (starvation) and voltage sharply increases, as seen in experiments at 0.22 ml/min and 10 mA as well as 0.44 ml/min and 20 mA. At high enough flow rates, however, voltage profiles collapse to a single profile determined by the cell capacitance value and Faradaic losses.

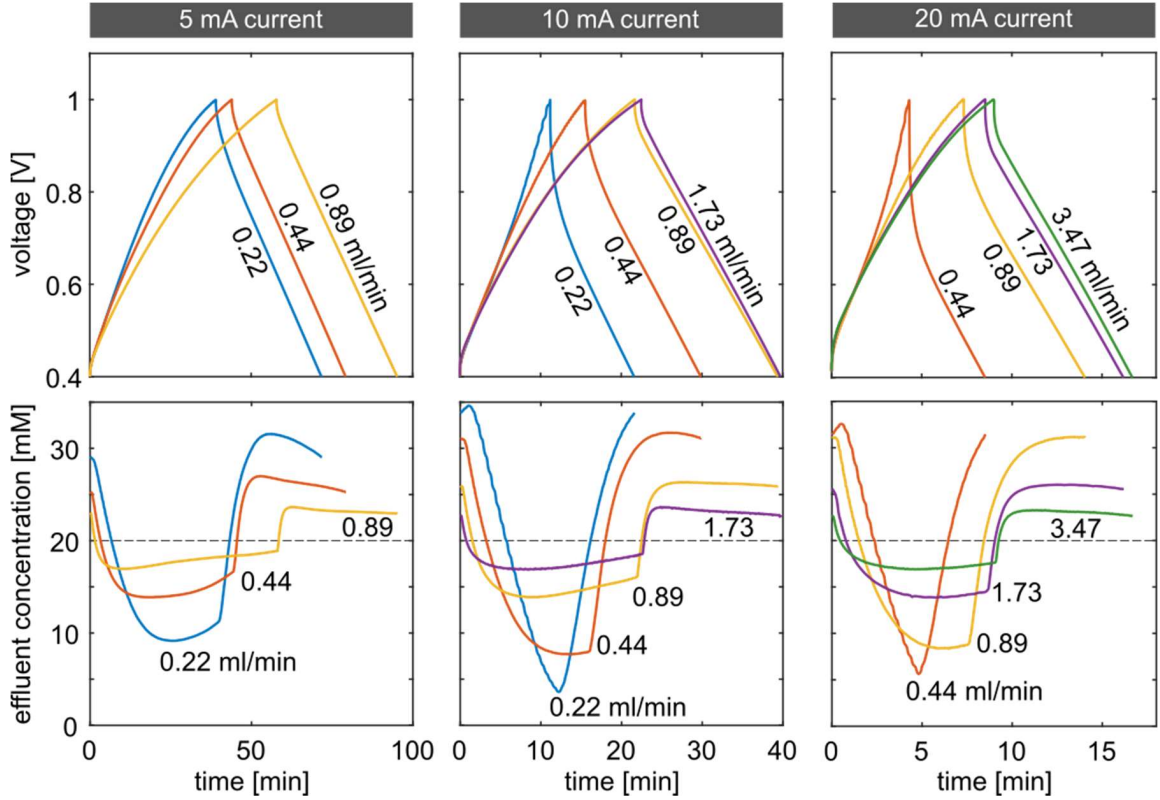

Figure 3. Selected voltage and effluent concentration profiles for constant current experiments with 20 mM NaCl inlet salt solution. Each column corresponds to experiments at a fixed current (5, 10, and 20 mA, respectively), and labels in each panel shows values of flow rate used.

## S.6 Cell performance data

We here discuss the variation of throughput, energy and desalination performance metrics with operational variables (flowrate and current). We define our metrics as follows. The average salt adsorption rate ($ASAR$) in units of moles of salt per total electrode area per time (or grams of salt per gram of electrode per time) is defined as

$$ASAR = \frac{Q}{NAt_{cycle}} \int_0^{t_{charge}} (c_0 - c) \mathrm{d}t, \tag{S.14}$$

where $N = 2$ is number for electrode pairs, $A = 25$ cm² is single electrode area, $t_{cycle}$ is cycle time, $t_{charge}$ is charging time, $Q$ is flow rate, and $c$ and $c_0$ are effluent and influent salt concentrations, respectively. *ASAR* is an indicator of throughput and average salt removal rate. The energy normalized adsorbed salt (*ENAS*) in units of moles of salt removed (or grams of salt removed) per Joules of energy lost is defined as[1]

$$ENAS = \frac{Q}{\Delta E} \int_0^{t_{charge}} (c_0 - c) dt, \tag{S.15}$$

where $\Delta E \ (= \int_0^{t_{cycle}} IV dt)$ represents the net energy input during one cycle. *ENAS* quantifies the energy efficiency of the desalination process. The average concentration reduction in the freshwater produced, $\Delta c$ in units of mmol is defined as

$$\Delta c = \frac{\int_0^{t_{charge}} Q(c_0 - c) dt}{\int_0^{t_{charge}} Q dt}. \tag{S.16}$$

Here, $\Delta c$ is a measure of desalination depth in the freshwater produced. The desalted water concentration is thus $c_D = c_0 - \Delta c$, corresponding to the desalinated volume $V_D = \int_0^{t_{charge}} Q dt$. The remaining feed water processed in a cycle results in concentrated brine solution with volume $V_B$ and concentration $c_B$. The recovery ratio is $\gamma = V_D/(V_B + V_D)$. Salt conservation requires that $c_0 = c_B(1-\gamma) + c_D \gamma$. Finally, the specific free energy of separation resulting from the desalination process $\Delta g_{sep}$ in units of Wh/m³ is given by

$$\Delta g_{sep} = 2RT \left[ \frac{c_0}{\gamma} \ln\left\{ \frac{c_0 - \gamma c_D}{c_0(1-\gamma)} \right\} - c_D \ln\left\{ \frac{c_0 - \gamma c_D}{c_D(1-\gamma)} \right\} \right], \tag{S.17}$$

where $\Delta g_{sep}$ represents the minimum energy required for the given desalination process.

Figure 4 shows the variation of CDI performance indicators versus operating conditions (current and flowrate). Figures 4a, 4c, and 4d show that at each flow rate, the three metrics: *ASAR*, $\Delta c$ and $\Delta g_{sep}$ increase, reach a maximum simultaneously and then decreases with increasing current. Further, note that the current corresponding to the maximum value in these metrics shifts to higher values for higher flow rates. Figure 4b shows that *ENAS* also increases, reaches a maximum and then decreases with increasing current. However, the current corresponding to peak *ENAS* is slightly lower than the current at which *ASAR*, $\Delta c$ and $\Delta g_{sep}$ are maximum. Also, from Figure 4a note that *ASAR* increases monotonically with increasing flowrate, whereas in Figures 4c and 4d, the peak values of $\Delta c$ and $\Delta g_{sep}$ (over varying current) decrease with increasing flow rate. More importantly, note in Figure 4b that the peak value of *ENAS* (over varying current) increases, reaches maximum and then decreases with increasing flowrate. This trend in *ENAS* is similar to the variation of thermodynamic efficiency with flowrate and current values in Figure 3 of the main text. Further, from Figure 4a, note that at low currents, *ASAR* for all flow rates asymptote to the same line (indicated by a dashed line). This is best explained using the following relation for a CC charging and discharging operation.

$$ASAR = \frac{I t_{charge} \Lambda_{cycle}}{FNA t_{cycle}}, \tag{S.18}$$

where $\Lambda_{cycle} = \bar{\lambda}_{EDL} \bar{\lambda}_c \lambda_{fl}$.[4] Here, $\bar{\lambda}_c = t_{discharge}/t_{charg}$ is a cycle average Coulombic efficiency, $\bar{\lambda}_{EDL}$ represents a cycle average dynamic EDL efficiency which is mainly a function of the voltage window, and $\lambda_{fl}$ is the flow efficiency which is a measure of salt recovery at the effluent. We note that in the limit of vanishing current $I \to 0$, we have $\lambda_{fl} \to 1$ for any finite flow rate.[4] Hence, in the limit $I \to 0$, Equation S.13 simplifies to

$$ASAR = \frac{I\gamma \bar{\lambda}_{EDL} \bar{\lambda}_c}{F\,N\,A}. \tag{S.19}$$

Taking the derivative of $ASAR$ with respect to current in this limit, we get

$$\frac{d(ASAR)}{dI} = \frac{\gamma \bar{\lambda}_{EDL} \bar{\lambda}_c}{F\,N\,A}. \tag{S.20}$$

where the right-hand side (RHS) of Equation S.20 is a constant value which determines the slope of the asymptotic line shown in Figure 4a. The slope (given by the RHS of Equation S.20) depends on recovery ratio, and average EDL and Coulombic efficiencies for the given voltage window.

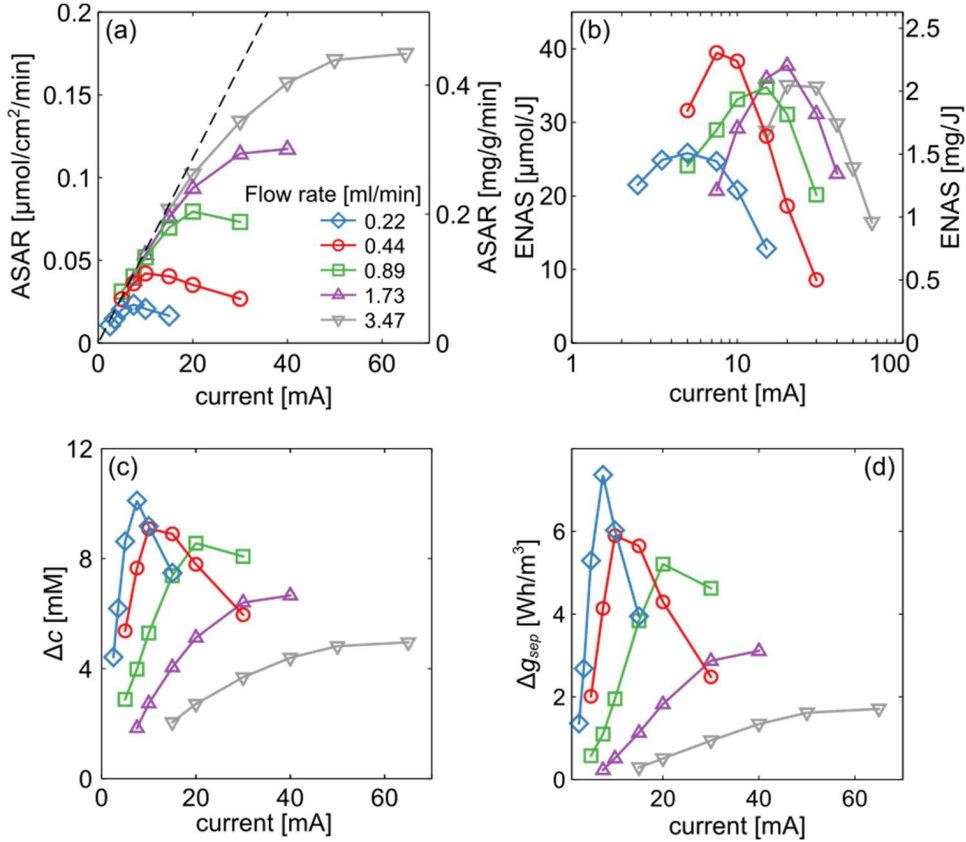

Figure 4. Measured values of CDI performance indicators: (a) Average salt adsorption rate ($ASAR$), (b) Energy normalized adsorbed salt ($ENAS$), (c) Average concentration reductino ($\Delta c$), and (d) Gibbs free energy of separation ($\Delta g_{sep}$), versus operating current (CC operation) for flow rates of 0.22, 0.44, 0.89, 1.73 and 3.47 ml/min. In all cases, the cell voltage window was 0.4-1 V. Dashed line in Fig. 4a represents an asymptotic variation of $ASAR$ with current for all flow rates.

## S.7 Free energy of separation

We here study the dependence of the thermodynamic minimum energy of ion separation (as given by the Gibbs free energy of separation, $\Delta g$) as a function of the desalination depth given by $\Delta c$ and recovery ratio $\gamma$. We show below that, to a good approximation, $\Delta g/c_0 \sim (\Delta c/c_0)^n$, where $a$ and $n$ depend only on the recovery ratio. Note, the Gibbs free energy per unit of fresh volume produced is given by

$$\Delta g = 2RT \left[ c_D \ln c_D + \left(\frac{1}{\gamma} - 1\right) c_B \ln c_B - \frac{c_0}{\gamma} \ln c_0 \right]. \quad \text{(S.21)}$$

Further, using mass conservation, we have

$$c_0 = c_B(1 - \gamma) + c_D \gamma \quad \text{(S.22)}$$

where, $c_D = c_0 - \Delta c$. Substituting Eqn. S.15 in S.14, and using a Taylor series expansion of S.14 about $\Delta c/c_0 = 0$, we obtain

$$\frac{\Delta g}{c_0} = 2RT \left[ \frac{1}{2(1-\gamma)} \left(\frac{\Delta c}{c_0}\right)^2 + \frac{1-2\gamma}{6(1-\gamma)^2} \left(\frac{\Delta c}{c_0}\right)^3 + O(\Delta c^4) \right] \quad \text{(S.23)}$$

Note that the leading order term in S.16 is quadratic in $\Delta c/c_0$. Hence, we expect that $\Delta g/c_0$ has a power law dependence with $\Delta c/c_0$, where the exponent is close to 2, i.e., we hypothesize,

$$\frac{\Delta g}{c_0} \approx a \left(\frac{\Delta c}{c_0}\right)^n \quad \text{(S.24)}$$

Figure 5a shows the variation of $\Delta g/c_0$ vs. $\Delta c/c_0$ for recovery ratios $\gamma$ of 0.25, 0.5, and 0.75, as given by Equation S.14 and from a power law fit, $\Delta g/c_0 = a(\Delta c/c_0)^n$. Note that the power law fits exceedingly well to the thermodynamic value of $\Delta g$ given by Equation S.14. We show the values of the fitting parameters $a$ and $n$ versus the recovery ratio in Figure 5b. Note that these are solely dependent on the recovery ratio, and that the exponent $n$ is close to 2 (between 1.5-2.5) for all recovery ratios. We note that for significant desalination depths $\Delta c/c_0 \to 1$, higher order terms are required to approximate the Taylor series expansion about $\Delta c/c_0 = 0$ as given in Equation S.16. Alternately, for such conditions, a Taylor series expansion around $\Delta c/c_0 = 1$ can be performed (not shown here), which results in a similar power law dependence. Finally, combining the power law (Equation S.17) with the resistive electrical work $E_{v,resist}$, we get,

$$\eta \sim \frac{\Delta g}{E_{v,resist}} \sim \frac{\gamma \Lambda_{cycle} a}{FIR_s} \left(\frac{\Delta c}{c_0}\right)^{n-1}. \quad \text{(S.25)}$$

In Equation S.18, we have used the relation $\Delta c = I\Lambda_{cycle}/FQ$, where $\Lambda_{cycle}$ is the cycle charge efficiency. Also, note that the exponent $n - 1$ is positive (close to 1). Thus, for a given recovery ratio (i.e., fixed $\gamma, a$ and $n$), thermodynamic efficiency of a desalination cycle can be increased by (*i*) increasing cycle charge efficiency, (*ii*) decreasing the series voltage drop $IR_s$ and, (*iii*) larger desalination depths, $\Delta c/c_0$.

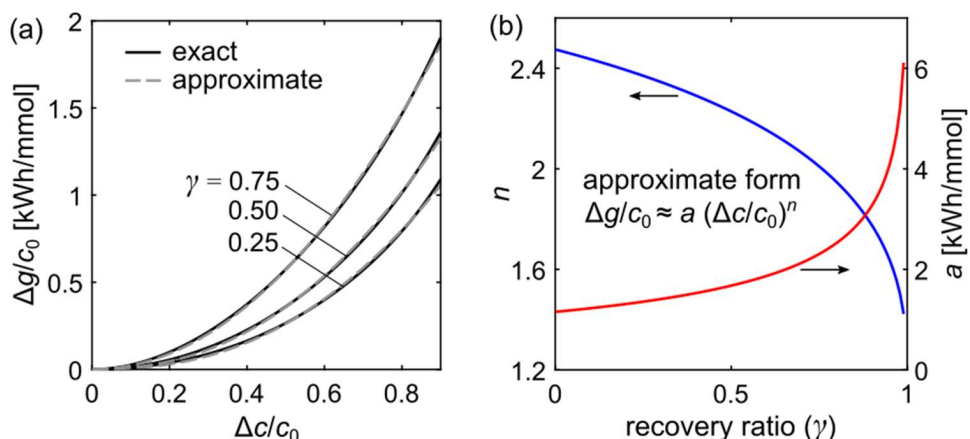

Figure 5. (a) Variation of the Gibbs free energy of separation normalized by feed concentration, $\Delta g/c_0$ vs. concentration reduction normalized by feed concentration, $\Delta c/c_0$ for varying recovery ratio values $\gamma = 0.25, 0.5$, and $0.75$. Solid black lines represent the exact variation from thermodynamic principles (Equation S.14), and the dashed grey lines represent a curve fit from the approximation of the form, $\Delta g/c_0 = a(\Delta c/c_0)^n$, where $a$ and $n$ are fitting parameters. The fit parameters $a$ and $n$ depend solely on the recovery ratio, and their variation is shown in (b).

### S.8 Practical considerations for efficient CC operation

For a cell with fixed system properties, we here delineate practical considerations for CC charging and discharging operation design that can ensure good overall performance.

(a) *Choice of current and flow rate*:
The flow rate and $Q/I$ ratio can be determined based on throughput and salt removal ($\Delta c = I\Lambda_{cycle}/FQ$) requirements, respectively, assuming a high cycle charge efficiency.

(b) *Choice of voltage thresholds*:
The maximum voltage threshold (after accounting for series resistance drop) can be fixed based on a Faradaic threshold. A minimum voltage threshold can be chosen so that least 4-5 cell volumes are flowed during charging to ensure efficient recovery of fresh water at the effluent. A good choice of voltage thresholds results in a high cycle charge efficiency (close to unity).

(c) *Avoiding extremities*:
Extreme operations such as (*i*) very low operating currents wherein the operating current is comparable to the leakage current, (*ii*) very low flow rates wherein dispersion and mixing inefficiencies are important, (*iii*) very high operating voltages wherein Faradaic losses dominate, can all adversely affect overall performance.

Any practical desalination operation is always a tradeoff among throughput, salt removal and energy efficiency. The considerations (a)-(c) listed above enable operation with good overall performance in all metrics.